\documentclass[twocolumn]{aastex6}

\usepackage{amssymb}
\begin{document}

\title{Energy spectra of abundant cosmic ray nuclei in sources, according to the ATIC experiment}

\shorttitle{Nuclei in sources, according to ATIC} 

\author{A.D. Panov\altaffilmark{1} and N.V. Sokolskaya and V.I. Zatsepin}
\affil{Moscow State University, Skobeltsyn Institute of Nuclear Physics, Moscow, 119991, Russia}
\altaffiltext{1}{panov@dec1.sinp.msu.ru}

\begin{abstract}
%% Text of abstract 
One of the main results of the  ATIC (Advanced Thin Ionization Calorimeter) experiment is a collection of energy spectra of abundant cosmic-ray nuclei: protons, He, C, O, Ne, Mg, Si, Fe measured in terms of energy per particle in the energy range from 50 GeV to tens of teraelectronvolts. In this paper, the ATIC energy spectra of abundant primary nuclei are back-propagated to the spectra in sources in terms of magnetic rigidity using a leaky-box approximation of three different GALPROP-based diffusion models of propagation that fit the latest B/C data of the AMS-02 experiment. It is shown that the results of a comparison of the slopes of the spectra in sources are weakly model dependent; therefore the differences of spectral indices are reliable data. A regular growth of the steepness of spectra in sources in the range of magnetic rigidity of 50--1350~GV is found for a charge range from helium to iron. This conclusion is statistically reliable with significance better than 3.2 standard deviations. The results are discussed and compared to the data of other modern experiments.
\end{abstract}


\section{Introduction}
\label{1}
The ATIC (Advanced Thin Ionization Calorimeter) balloon spectrometer was designed to measure the energy spectra of primary cosmic-ray nuclei from protons to iron with elemental charge resolution in the energy range of $\sim$50 GeV to 100~TeV per particle \citep{ATIC-2004-GUZIK-AdvSpRes}. It was shown that the spectrometer is also capable of measuring the total spectrum of cosmic-ray electrons and positrons \citep{ATIC-2008-CHANG-NATURE}. ATIC had three successful flights around the South Pole: in 2000--2001 (ATIC-1), in 2002--2003 (ATIC-2), and in 2007--2008 (ATIC-4). ATIC-1 was a test flight; nuclear spectra from protons to iron and the spectrum of electrons were measured in the ATIC-2 flight; and the electron spectrum only was measured in ATIC-4, due to malfunctioning of the pretrigger system. The present work is based on the results of the ATIC-2 flight.

The ATIC spectrometer consists of a fully active BGO calorimeter, a carbon target with embedded scintillator hodoscopes, and a matrix of silicon detectors. The silicon matrix is used as a primary particle charge detector. The design of the instrument and the calibration procedures were described in detail elsewhere 
\citep{ATIC-2004-GUZIK-AdvSpRes,ATIC-2004-ZATSEPIN-NIM,ATIC-2008-PANOV-IET-ENG}.

% \section{Rigidity spectra of abundant nuclei observed near the Earth}
% \label{2}

\begin{figure}
\includegraphics[width=0.45\textwidth]{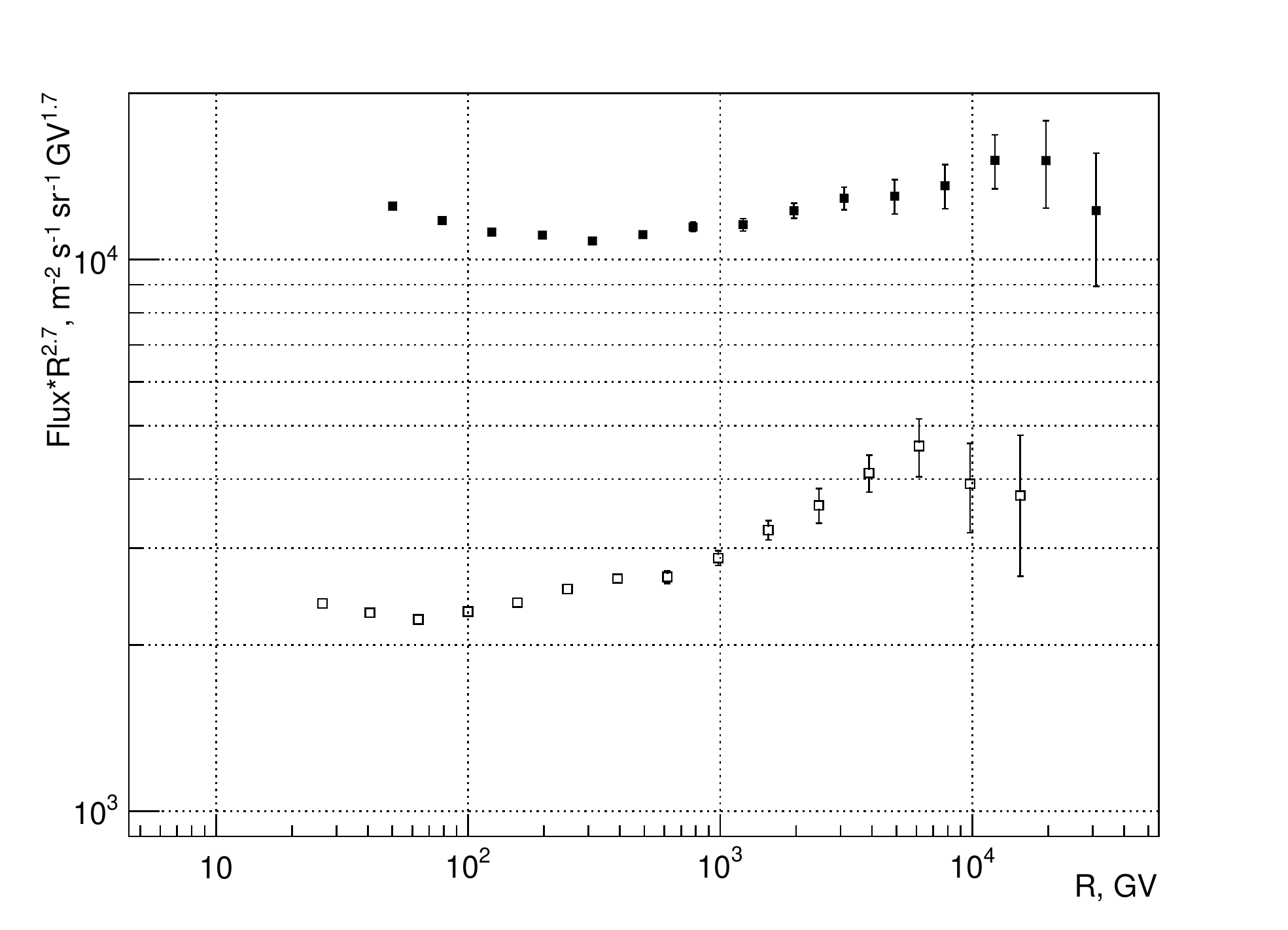}
\caption{\label{pHe-Rig} Rigidity spectra of protons (solid squares) and helium (open squares) near the Earth measured by ATIC.}
\end{figure}

\begin{figure}
\includegraphics[width=0.45\textwidth]{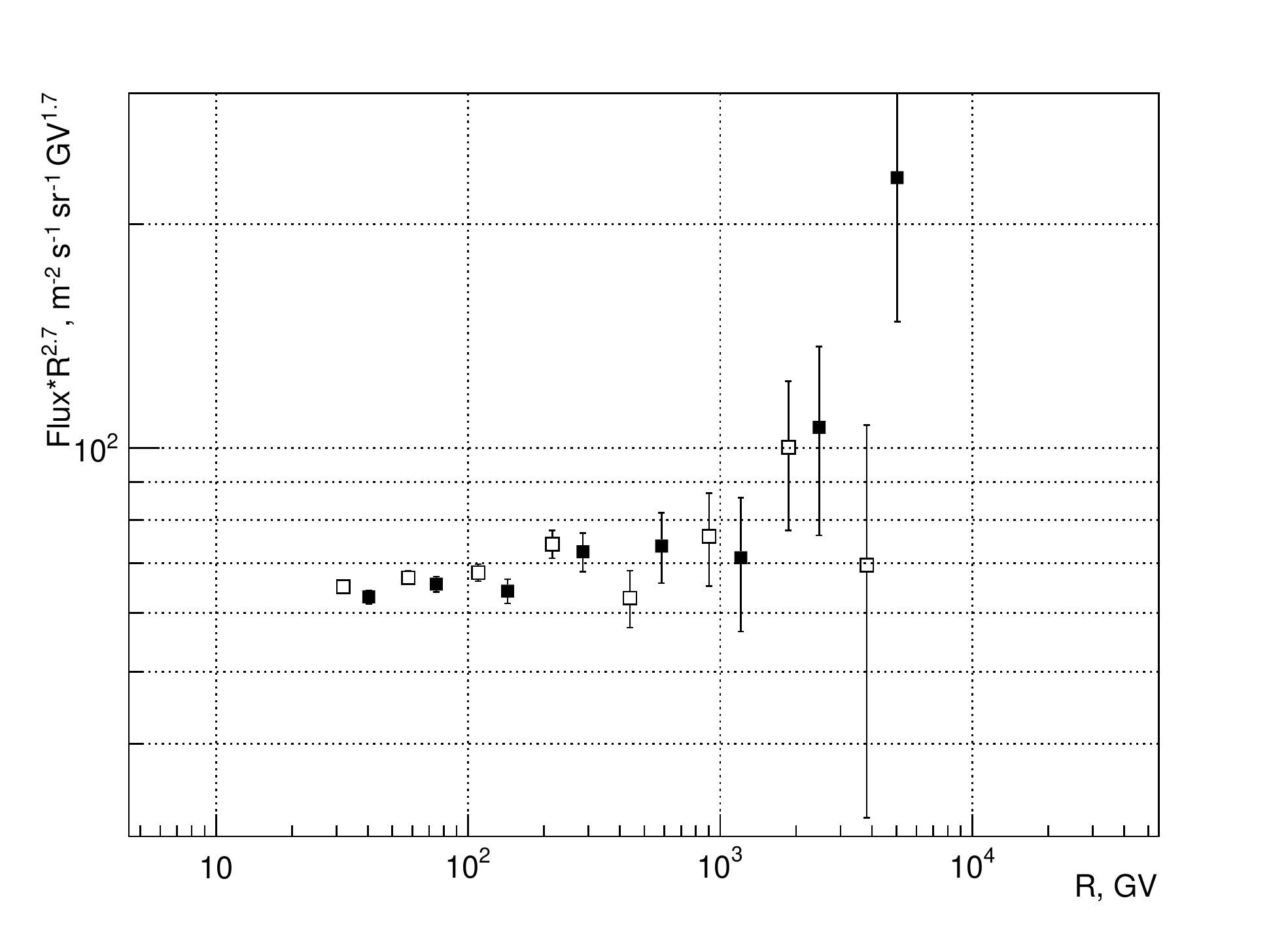}
\caption{\label{CandO} Rigidity spectra of carbon (solid squares) and oxygen (open squares) near the Earth measured by ATIC.}
\end{figure}

\begin{figure}
\includegraphics[width=0.45\textwidth]{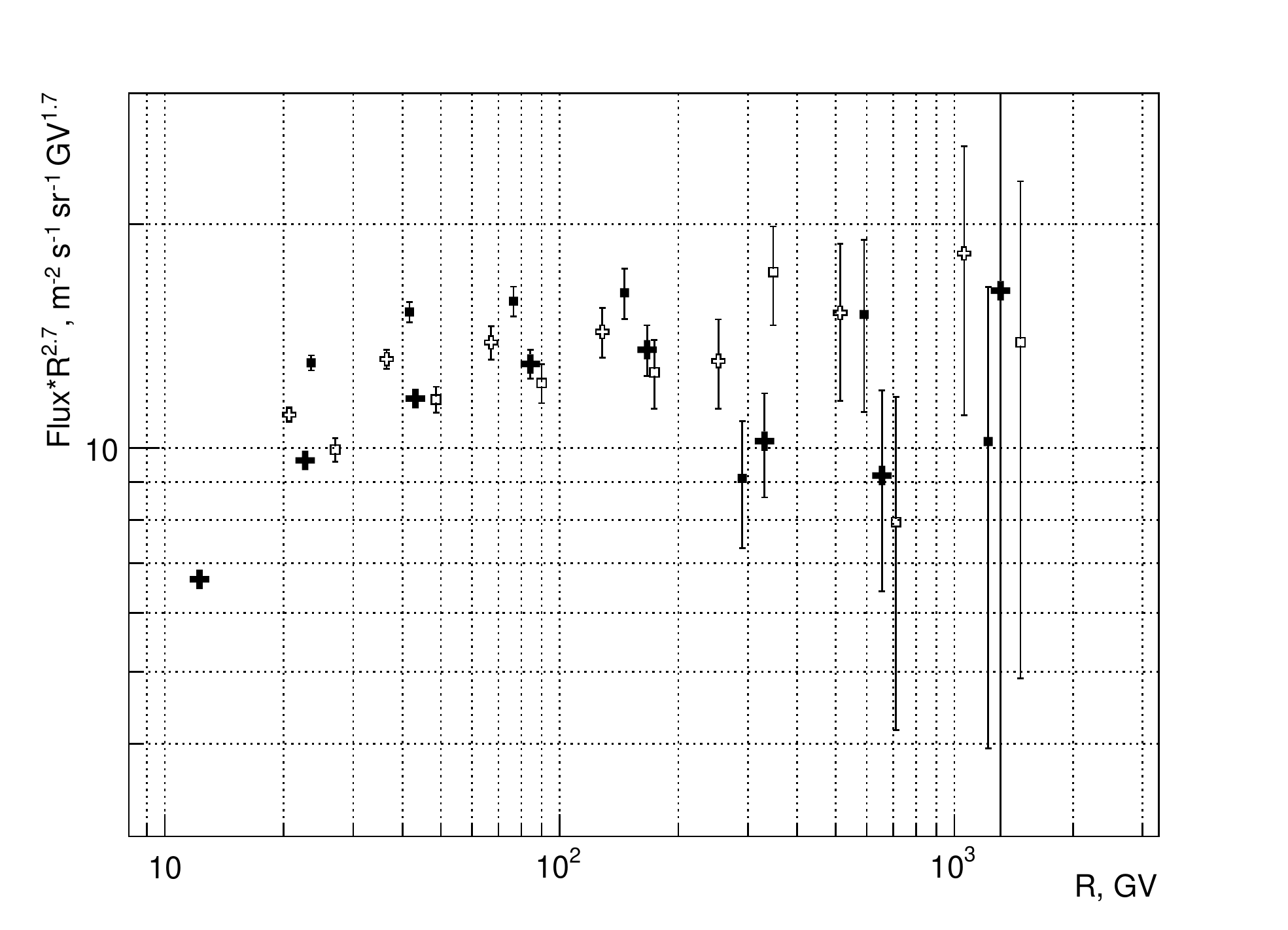}
\caption{\label{NeMgSiFe} Rigidity spectra of Ne (open squares), Mg (solid squares), Si (open crosses) and Fe (solid crosses) near the Earth measured by ATIC.}
\end{figure}

The data obtained by the ATIC spectrometer include high-precision energy spectra of the most abundant cosmic-ray nuclei (protons, He, C, O, Ne, Mg, Si, and Fe) measured in terms of energy per particle \citep{ATIC-2009-PANOV-IzvRAN-ENG}. The total kinetic energy per particle is the most natural quantity for expressing the energy measured by the calorimetric spectrometer, and the results of the ATIC experiment were given in this way in \citep{ATIC-2009-PANOV-IzvRAN-ENG}. From the viewpoint of the physics of the propagation and acceleration of cosmic rays, however, it is more important to know the magnetic rigidity spectra of cosmic rays, and it is the information on the rigidity spectra \emph{in sources} that is most important in studying the mechanisms of the acceleration of cosmic rays. Converting from the observed energy per particle spectra to the rigidity spectra using the ATIC data poses no difficulties, since the charge of each particle is measured along with its energy. The rigidity spectra of abundant nuclei measured by the ATIC spectrometer are shown in Figs.~\ref{pHe-Rig}--\ref{NeMgSiFe}. They are obtained directly from the energy per particle spectra published in \citep{ATIC-2009-PANOV-IzvRAN-ENG}. Our objective in the present paper is to obtain and compare the spectra of abundant primary nuclei from protons to iron in the source, which have not hitherto been reported. From the point of view of the physics of propagation and acceleration of cosmic rays such a comparison is very important.

\section{Solving the inverse propagation problem in the leaky-box approximation}
\label{3}

In order to obtain the source spectra in terms of magnetic rigidity, the inverse problem of propagation of cosmic-ray particles must be solved using one model of propagation or another. We restrict ourselves in this paper to propagation models that consider the interstellar medium to be smooth within the magnetic galactic halo. Some more complicated models that consider the interstellar medium to be essentially inhomogeneous have been studied elsewhere (see, for example, the model of a closed galaxy with super bubbles embedded; \citep{CRPROP-PETERS1977,ATIC-2014-NuclPhysB}). However, at the present time the basis of such models looks insufficiently firm. For example, the model of Local Bubble within closed galaxy explains the upturn in the ratio of fluxes of nuclei of Z=16-24 to iron near the energy 50~GeV/n  \citep{ATIC-2014-NuclPhysB}, but the same model is in contradiction with the latest data on B/C ratio of PAMELA \citep{CR-PAMELA-2014-ApJ-BtoC} and AMS-02 \citep{AMS-02-2016-PRL-BtoC} showing no such upturn. Therefore, we will consider only the simplest case of a smooth interstellar medium.

It is generally quite difficult to solve the inverse problem for the general diffusion transport equation. However, the problem can be essentially simplified within a homogeneous model \citep{COWSIK1967-PhysRev,CRPROP-PTUSKIN1975-ENG,GAISSER1990} known as the leaky-box approximation to the diffusion transport equation. 

The leaky-box approximation was compared with more sophisticated diffusion models in \citep{CRPROP-LETAW1993,CRPROP-PTUSKIN2009}. The leaky-box approximation works well for stable and not-too-heavy cosmic-ray nuclei. With a number of different assumptions about the character of diffusion, it was shown in \citep{CRPROP-PTUSKIN2009} that numerical solutions of the diffusion equation for fluxes of stable nuclei obtained with the GALPROP system could be approximated within a percentage accuracy using leaky-box models for nuclei from protons up to nickel. That is, the exact solution of the diffusion equation yields essentially the same results as a properly constructed leaky-box model, so one can use the leaky-box model instead of an explicit solution of the diffusion equation. In the leaky-box model, a solution of the inverse problem of propagation may be obtained easily.  

The diffusion of cosmic rays in the Galaxy is described in the leaky-box model by a single parameter: the particle diffusion escape length from the Galaxy $\lambda_{esc}(R)$, measured in g\,cm$^{-2}$, which depends only on the magnetic rigidity $R$ of the particles. If primary abundant nuclei of a certain type are described by an effective source averaged over the Galaxy volume with rigidity spectrum $Q(R)$, their observed equilibrium spectrum takes the following form:
\begin{equation}
\label{eq:LB-Direct}
M(R) = \frac{1}{\rho v}\frac{1}{[1/\lambda_{esc}(R) + 1/\lambda_N]}Q(R),
\end{equation}
where $\lambda_N$ (g\,cm$^{-2}$) is the mean free path of a nucleus before nuclear interaction with fragmentation in the interstellar medium, $v$ is the velocity of the particle, and $\rho$ is the interstellar medium density averaged by the volume of the magnetic halo. 

Equation (\ref{eq:LB-Direct}) is a solution of the direct problem of cosmic-ray propagation for the considered special case. The solution of the inverse problem of propagation (i.e., determining the source function from an observed particle spectrum) is obtained through trivial inversion of Eq.~(\ref{eq:LB-Direct}):
\begin{equation}
\label{eq:LB-Reverse}
Q(R) = \rho v [1/\lambda_{esc}(R) + 1/\lambda_N] M(R).
\end{equation}

Three different GALPROP models were studied in \citep{CRPROP-PTUSKIN2009} to support the conclusion that the leaky-box model approximates well the results of explicit solutions of the diffusion equation for the primary abundant cosmic-ray nuclei. The \emph{plain} GALPROP model had no reacceleration and was characterized by the following diffusion coefficient:
\begin{equation}
  \begin{array}{l}
  D(R) = D_0(R/3\ \mathrm{GV})^\delta,\nonumber\\
  D_0  = 2.2\times10^{28}\mathrm{cm}^2/\mathrm{s},\ \delta = 0.6.
  \end{array}
  \label{D0Plain}
\end{equation}
We neglected by the factor $\beta = v/c$ here since $\beta \simeq 1$ in the range of our interest.
The GALPROP model with \emph{distributed reacceleration} in a medium with Kolmogorov turbulence had the following parameters:
\begin{eqnarray}
\begin{array}{l}
 D_0 = 5.2\times 10^{28}\mathrm{cm}^2/\mathrm{s},\ \delta = 0.34\\
 V_a = 36\ \mathrm{km/s},
\end{array}
 \label{D0Reaccel}
\end{eqnarray}
where  $V_a$ is the Alfven velocity, and $D(R)$ is again a power-law function like in Eq.~(\ref{D0Plain}). Finally, the  model \emph{with damping} was characterized by a mechanism of the back reaction of cosmic rays on the interstellar Kraichnan-type turbulence together with distributed reacceleration. The following set of parameters was used in the damping model:
\begin{eqnarray}
\begin{array}{l}
 D_0 = 2.9\times 10^{28}\mathrm{cm}^2/\mathrm{s},\ \delta = 0.5\\
 V_a = 22\ \mathrm{km/s},
\end{array}
\label{D0Damping}
\end{eqnarray}
The half-width of the galactic magnetic halo was $H = 4$\,kpc in all of these models. We will briefly refer to these models as the plain Equation~(\ref{D0Plain}), the reacceleration Equation~(\ref{D0Reaccel}), and the damping  Equation~(\ref{D0Damping}).  Some further details of the models may be found in \citep{CRPROP-PTUSKIN2009}. 

\begin{figure}
\includegraphics[width=0.45\textwidth]{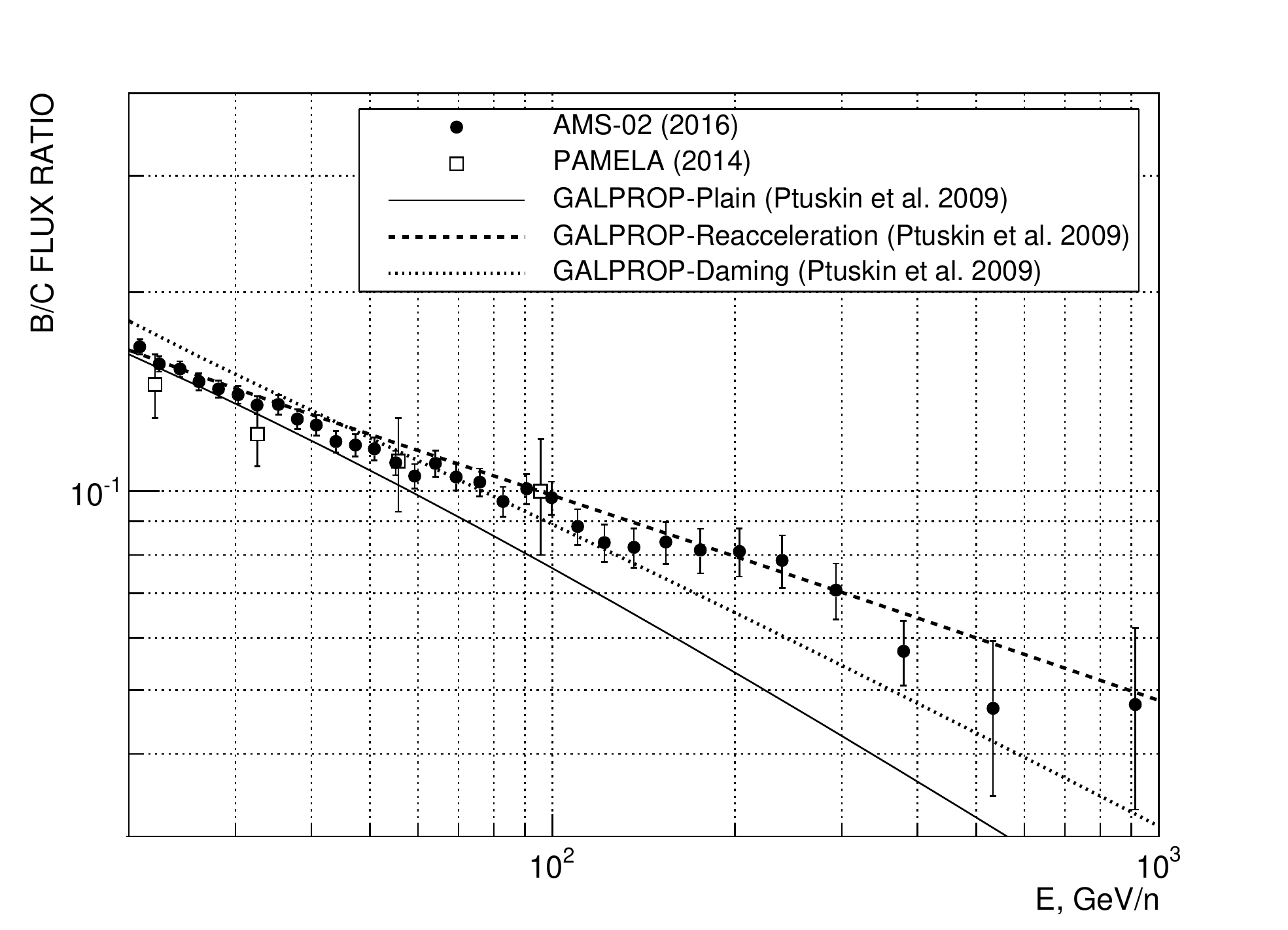}
\caption{\label{AMS-PAMELA-GALPROP} Boron to carbon ratio by PAMELA \citep{CR-PAMELA-2014-ApJ-BtoC} and AMS-02 \citep{AMS-02-2016-PRL-BtoC} and the original GALPROP models \citep{CRPROP-PTUSKIN2009} prediction.}
\end{figure}

The prediction of the models of Equations (\ref{D0Plain})--(\ref{D0Damping}) for the boron-to-carbon ratio obtained with the GALPROP system \citep{GALPROP-Web-2011} is shown in Figure~\ref{AMS-PAMELA-GALPROP} together with the latest data of PAMELA \citep{CR-PAMELA-2014-ApJ-BtoC} and AMS-02 \citep{AMS-02-2016-PRL-BtoC}. The data are shown for the rigidities $R > 20$\,GV relevant to the present paper. It is seen from Figure~\ref{AMS-PAMELA-GALPROP} that the original models of Equations(\ref{D0Plain})--(\ref{D0Damping}) do not fit the data of AMS-02 well. The data of the PAMELA experiment have too few points above $R=20$\,GV with rather high statistical errors, so we will use the high-precision data of AMS-02 only in our analysis. The AMS-02 data supply a much more firm basis for quantitative analysis and were used in the paper. However, we would like to note that the PAMELA data together with the most recent AMS-02 data show that there is no great systematic uncertainty in the contemporary experimental B/C ratio.

\begin{figure}
\includegraphics[width=0.45\textwidth]{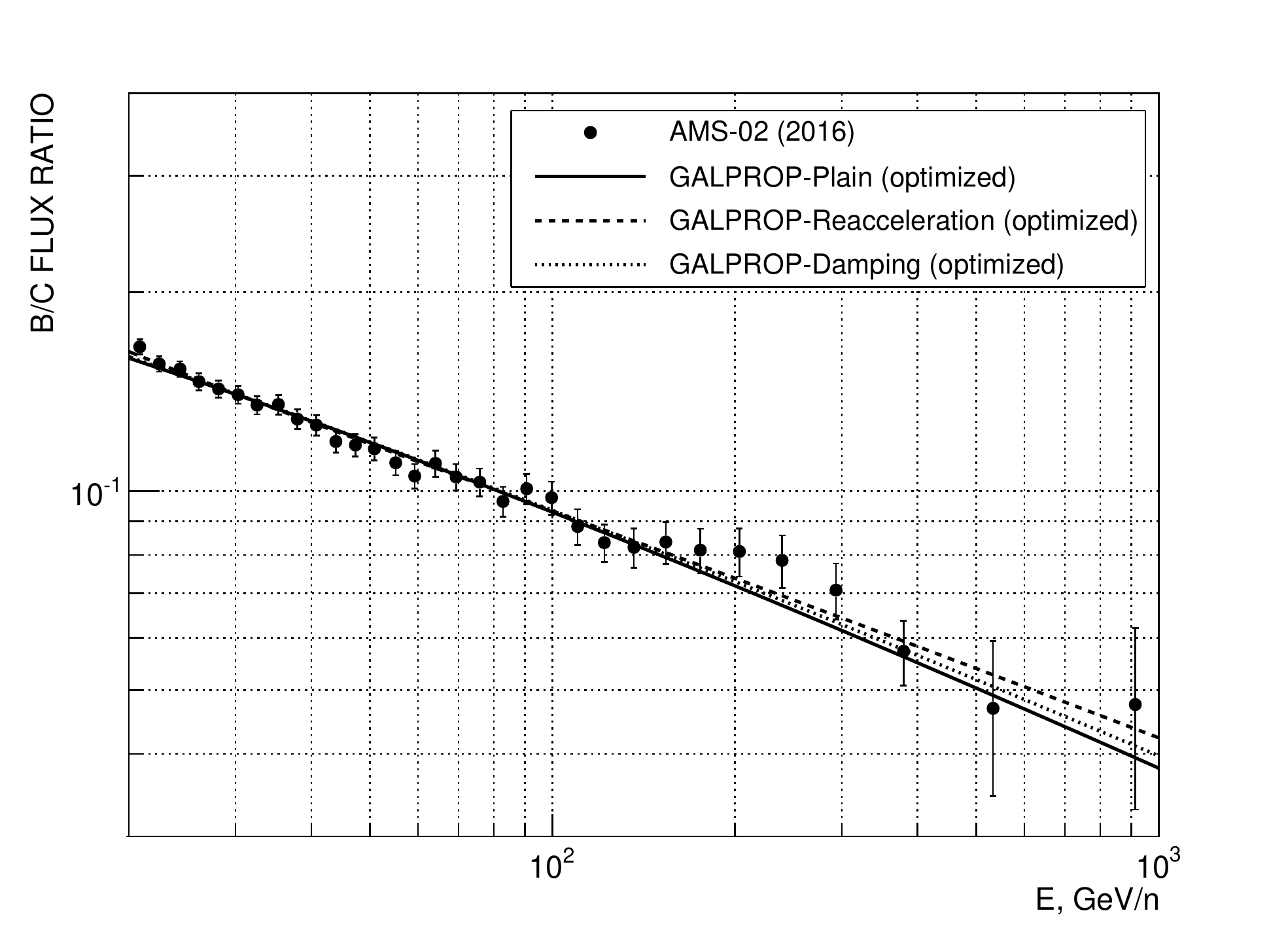}
\caption{\label{AMS-OPTIMAL} Boron to carbon ratio by AMS-02 \citep{AMS-02-2016-PRL-BtoC} and the optimized plain, reacceleration and damping models.}
\end{figure}

The same models of Equations(\ref{D0Plain})--(\ref{D0Damping}) may be fitted to the AMS-02 data optimally by a variation of the parameters $D_0$ and $\delta$ of the power-law function $D(R)$. The results of such an optimization are shown in Figure~\ref{AMS-OPTIMAL}. The optimization hase been carried out for the rigidity region $R > 20$\,GV. It is seen that the optimized curves for different models are a bit different, and the best $\chi^2$ values per unit of freedom are different for different models as well: 0.668, 0.437, and 0.596 for the plain, reacceleration, and damping models, respectively. The reacceleration model fits the data better than the two other models do. Also, different models produced the following different sets of parameters:
\begin{equation}
 \begin{array}{l}
  D_0 = (3.48 \pm 0.05)\times10^{28}\mathrm{cm}^2/\mathrm{s}\\
  \delta = 0.439 \pm 0.019\\
  \mathrm{corr} = -0.996
 \end{array}
  \label{DOptPlain}
\end{equation}
for the plain model;
\begin{equation}
 \begin{array}{l}
  D_0 = (4.74 \pm 0.06)\times10^{28}\mathrm{cm}^2/\mathrm{s}\\
  \delta = 0.375 \pm 0.037\\
  \mathrm{corr} = -0.996
 \end{array}
  \label{DOptReaccel}
\end{equation}
for the reacceleration model;
\begin{equation}
 \begin{array}{l}
  D_0 = (4.88 \pm 0.08)\times10^{28}\mathrm{cm}^2/\mathrm{s}\\
  \delta = 0.418 \pm 0.049\\
  \mathrm{corr} = -0.986
 \end{array}
  \label{DOptDamping}
\end{equation}
for the damping model. Here the errors are the statistical ones and mean one standard deviation; '$\mathrm{corr}$' means correlation of the statistical deviation for the parameters $D_0$ and $\delta$ (there is a strong anticorrelation). 

The starting point to obtain the leaky-box parameters to solve the back-propagation problem in our approach is the diffusion coefficients of Equations (\ref{DOptPlain})--(\ref{DOptDamping}). To obtain the leaky-box parameters to approximate the solution of diffusion equations with a given diffusion coefficient in a given model, the following method was used in \citep{CRPROP-PTUSKIN2009}. The quantity named effective escape length $X_{\mathrm{eff}}$ has been introduced. For a nucleus with the nuclear fragmentation length $\lambda_N$, $X_{eff}$ can be written as (see Eq.~(18) in \citep{CRPROP-PTUSKIN2009}):
\begin{equation}
 X_{\mathrm{eff}}(R) = \lambda_N\left[\frac{M_\infty(R)}{M_{\lambda_N}(R)} - 1\right],
 \label{Xeff}
\end{equation}
where $M_\infty$ is the computed flux of the nucleus supposing no fragmentation, and $M_{\lambda_N}$ is the computed flux for the actual value of $\lambda_N$. It is easy to see that $X_{\mathrm{eff}}(R)$ is exactly $\lambda_{esc}(R)$ within the leaky-box approximation. To obtain a leaky-box approximation for a diffusion \mbox{GALPROP} model, the effective escape length of Equation~(\ref{Xeff}), computed with this model, should be approximated by a power-law function $\lambda_{esc}(R) = \lambda_0 (R/R_0)^{-\delta}$, where $R_0$ is some arbitrary reference value of rigidity ($R_0= 3$\,GV in \citep{CRPROP-PTUSKIN2009}). The escape length parameters found in \citep{CRPROP-PTUSKIN2009}) this way were
\begin{equation}
 \lambda_0 = 19 \mathrm{g/cm}^2,\ \delta = 0.6
 \label{PtusL0deltaPlain}
\end{equation}
for the plain model;
\begin{equation}
 \lambda_0 = 7.2 \mathrm{g/cm}^2,\ \delta = 0.34
 \label{PtusL0deltaReaccel}
\end{equation}
for the reacceleration model;
\begin{equation}
 \lambda_0 = 13 \mathrm{g/cm}^2,\ \delta = 0.5
 \label{PtusL0deltaDamping}
\end{equation}
for the damping model.

To obtain the escape length of the leaky-box model for the optimized diffusion coefficients of Equations (\ref{DOptPlain})--(\ref{DOptDamping}) we do not reproduce the calculations of \citep{CRPROP-PTUSKIN2009} for each of the diffusion coefficients of Equations (\ref{DOptPlain})--(\ref{DOptDamping}), but use the results of \citep{CRPROP-PTUSKIN2009} directly in a more simple way. Starting from the usual one-dimensional formula for the diffusion length $\langle x \rangle = \sqrt{2D\tau}$, it is not difficult to show \citep{GAISSER1990} that within the leaky-box approximation and in the flat magnetic halo model there is the following relation between the escape length and the diffusion coefficient:
\begin{equation}
 \lambda_{esc} = \frac{1}{2}c\rho_d h_d\frac{H}{D}  \equiv K_d\frac{H}{D},
 \label{LamEscVsD}
\end{equation}
where $\rho_d$ is some effective mean density of the interstellar gas within the galaxy disk, $h_d$ is the effective half-width of the galaxy disk, and $H$ is the half-width of the magnetic halo. It is seen from Eq.~(\ref{LamEscVsD}) that if the exponent of a power-law diffusion coefficient is $\delta$, then the escape length is a power-law function as well, and its exponent is $-\delta$. If the diffusion coefficient of some diffusion model and the approximation of this diffusion model by a leaky-box model is known, then the disk factor $K_d$ for this pair of models could be easily obtained from Equation~(\ref{LamEscVsD}):
\begin{equation}
 K_d = \lambda_{esc}\frac{D}{H}
\end{equation}
If the coefficient $K_d$ for some specific diffusion model is known, then Eq.~(\ref{LamEscVsD}) can be used to calculate $\lambda_{esc}$ for any definite value of $D$.

\begin{figure}
\includegraphics[width=0.45\textwidth]{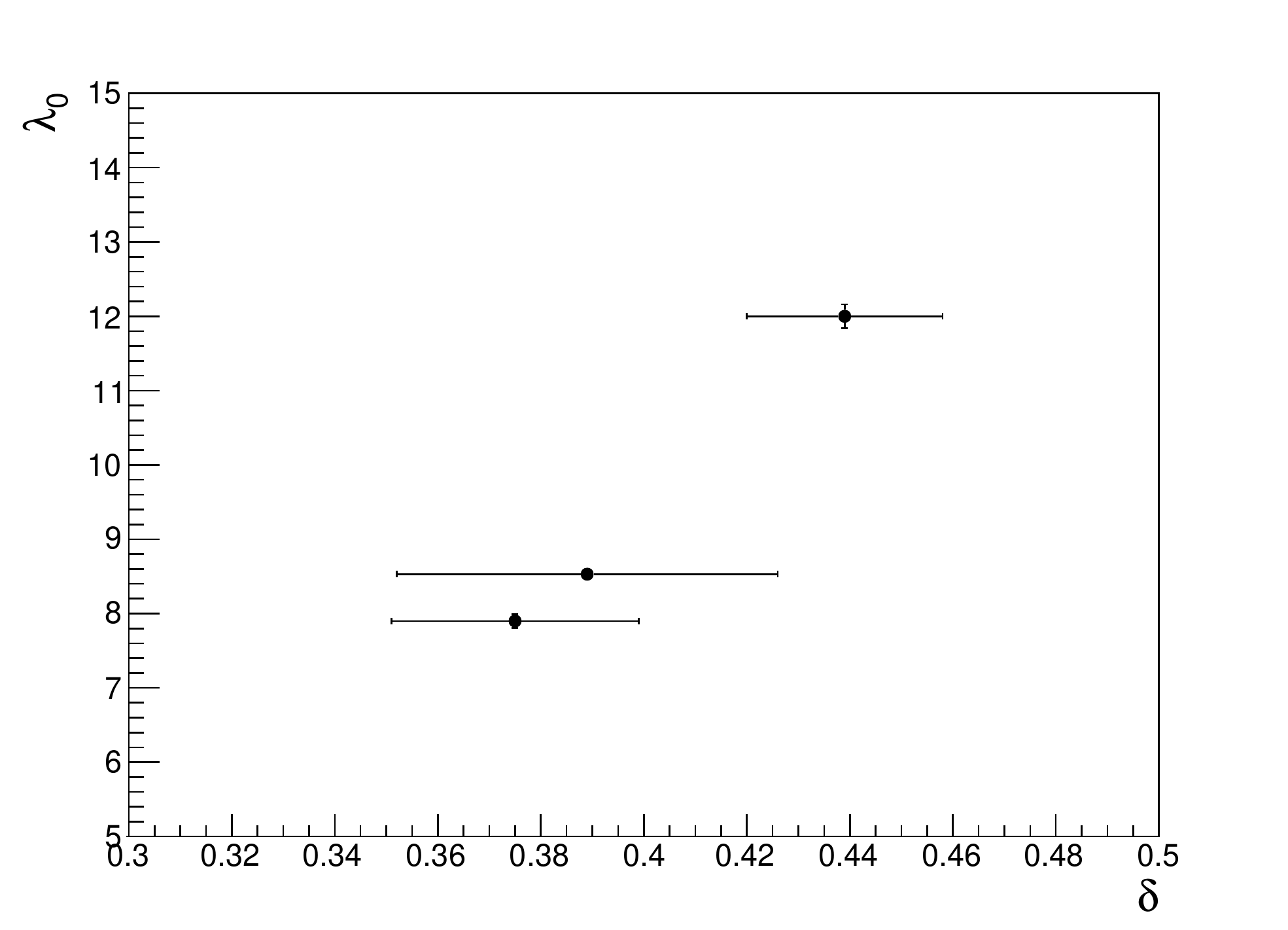}
\caption{\label{DELTA-LAMBDA} The results of approximation of three optimized GALPROP models (\ref{DOptPlain})--(\ref{DOptDamping}), see also Figure~\ref{AMS-OPTIMAL}, by the leaky-box approximations (\ref{OptLam0deltaPlain})--(\ref{OptLam0deltaDamping}).}
\end{figure}

It is expected generally that the coefficients $K_d$ calculated within different diffusion models should be close to each other since they refer to the same galaxy disk structure. Actually, using the parameters of Equations (\ref{D0Plain})--(\ref{D0Damping}) and (\ref{PtusL0deltaPlain})--(\ref{PtusL0deltaDamping}) for the original models of \citep{CRPROP-PTUSKIN2009} one obtains
\begin{eqnarray}
 K_d(\mathrm{plain})          &=& 1.045\times10^{29}\, \mathrm{g}\cdot\mathrm{s}^{-1}\!\cdot\mathrm{kpc}^{-1} \label{K_d-plain}        \\
 K_d(\mathrm{reacceleration}) &=& 0.936\times10^{29}\, \mathrm{g}\cdot\mathrm{s}^{-1}\!\cdot\mathrm{kpc}^{-1} \label{K_d-reacceleration}\\
 K_d(\mathrm{damping})        &=& 0.943\times10^{29}\, \mathrm{g}\cdot\mathrm{s}^{-1}\!\cdot\mathrm{kpc}^{-1}.\label{K_d-damping}
\end{eqnarray}
The values $K_d$ are the same within $10\%$ accuracy, and the difference is an expression of an approximate character of the leaky-box model. Now the values of Equations (\ref{K_d-plain})--(\ref{K_d-damping}) can be used to approximate the plain, reacceleration, and damping diffusion models with the optimized diffusion coefficients of Equations(\ref{DOptPlain}), (\ref{DOptReaccel}), and (\ref{DOptDamping}) respectively by the leaky-box models. The results are
\begin{equation}
 \lambda_0 = 12.0 \pm 0.2\, \mathrm{g}\cdot\mathrm{cm}^{-2},\ \delta = 0.439 \pm 0.019
 \label{OptLam0deltaPlain}
\end{equation}
\begin{equation}
 \lambda_0 = 7.90 \pm 0.09\, \mathrm{g}\cdot\mathrm{cm}^{-2},\ \delta = 0.375 \pm 0.024
 \label{OptLam0deltaReaccel}
\end{equation}
\begin{equation}
 \lambda_0 = 8.53 \pm 0.09\, \mathrm{g}\cdot\mathrm{cm}^{-2},\ \delta = 0.389 \pm 0.037
 \label{OptLam0deltaDamping}
\end{equation}
for the plain, reacceleration, and damping models, respectively. All of the obtained points are shown in Figure~\ref{DELTA-LAMBDA} in the axes $(\delta,\lambda_0)$.

It is seen from equations (\ref{DOptPlain})--(\ref{DOptDamping}) and (\ref{OptLam0deltaPlain})--(\ref{OptLam0deltaDamping}) that different diffusion models produce rather different diffusion coefficients or escape lengths in the leaky-box approximation to fit the same experimental data (Figure~\ref{AMS-OPTIMAL}). Therefore, there is a great systematic uncertainty in obtaining the escape length related to uncertainty of the diffusion model of propagation of cosmic-ray particles. It is easy to obtain a source spectrum from the measured spectrum of a given primary abundant nucleus using Equation~(\ref{eq:LB-Reverse}) with some known escape length,  but it is a misleading goal to obtain the source spectra of the cosmic-ray nuclei themselves: the uncertainty of the slopes of the spectra will be too large due to a systematic uncertainty of the escape length. However, as shown below, the \emph{differences} of the slopes (and, more generally, the shapes) of the source spectra of different nuclei can be dealt with quite reasonably.
 
\section{Differences between the shapes of source spectra for different abundant nuclei}
\label{4}
To compare the shapes of the source spectra for different nuclei, one can use the ratios of the spectra: when the ratio is not constant, the shapes of the spectra are different. The question that we want to address is, what is the level of stability of such ratios against the uncertainties of the propagation model used?

To answer this question, we construct a kind of a robust estimate of the relevant systematic errors. Let $\lambda_r$ and $\delta_r$ be the mean values of all $\lambda_0$ and $\delta_0$, respectively, determined by the Equations (\ref{OptLam0deltaPlain})--(\ref{OptLam0deltaDamping}). Then we have \emph{the reference leaky-box model} defined by
\begin{equation}
\lambda_r \approx 9.5\,\mathrm{g}\,\mathrm{cm}^{-2},\ \delta_r \approx 0.4.  
\label{Ref}
\end{equation}
Let $(\delta_{max} - \delta_{min})/2 = \Delta\delta_s = 0.032$ be an estimate of the systematic uncertainty for the value $\delta_r$, and let $\sigma_{max}\delta = 0.037$ be the maximal statistical error of $\delta$ among all the values defined by Equations (\ref{OptLam0deltaPlain})--(\ref{OptLam0deltaDamping}). Then for a robust estimate of the possible systematic error for $\delta_r$, we adopt the \emph{double} quadratic sum of $\Delta\delta_s$ and $\sigma_{max}\delta$:
\begin{equation}
 \Delta\delta = 2\sqrt{(\Delta\delta_s)^2+(\sigma_{max}\delta)^2} \approx 0.1.
\end{equation}
In a similar way, we obtain a robust estimate of the possible systematic error for $\lambda_r$:
\begin{equation}
 \Delta\lambda = 2\sqrt{(\Delta\lambda_s)^2+(\sigma_{max}\lambda)^2} \approx 4.1\,\mathrm{g}\cdot\mathrm{cm}^2.
\end{equation}

To estimate the stability of the ratios of spectra in the source against systematic uncertainties, we calculate them for the reference propagation model $(\lambda_r,\delta_r)$ and compare the result with four types of deviations in the parameters of the model: 
$(\lambda_r + \Delta\lambda,\delta_r+\Delta\delta)$, $(\lambda_r - \Delta\lambda,\delta_r-\Delta\delta)$, $(\lambda_r - \Delta\lambda,\delta_r+\Delta\delta)$, and $(\lambda_r + \Delta\lambda,\delta_r-\Delta\delta)$.

\begin{figure}
\includegraphics[width=0.45\textwidth]{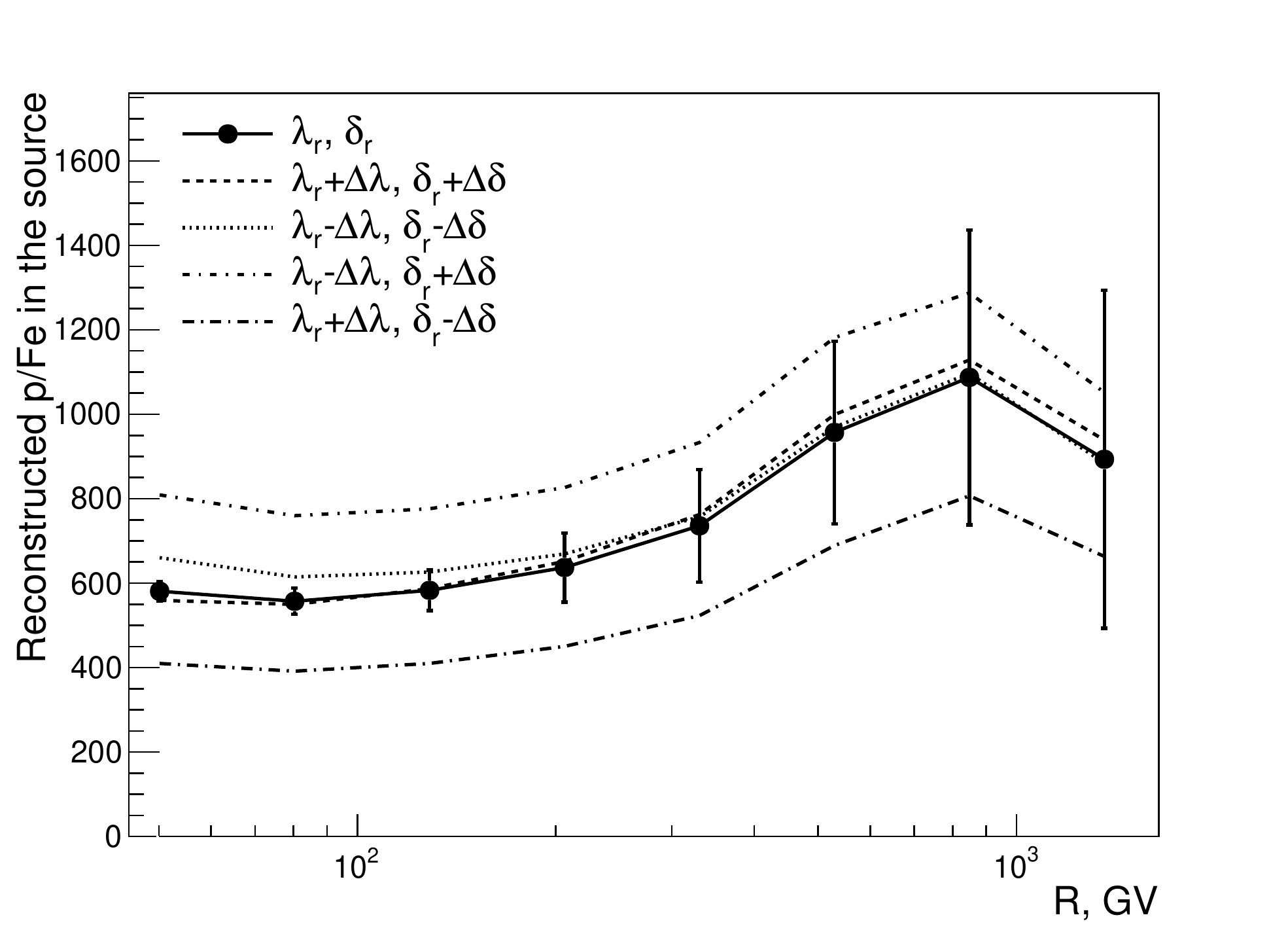}
\caption{\label{Ratio-p-Fe} Reconstructed ratio of the p/Fe source spectra, obtained with the reference propagation model (\ref{Ref}) and with the disturbed models. Statistical errors are specified only for the reference model since for all other models the relative errors are the same for each rigidity value $R$.}
\end{figure}

The ratios of spectra p/Fe in the source obtained with the reference model of Equation~(\ref{Ref}) and with four disturbed models are shown in Fig.~\ref{Ratio-p-Fe}. It is seen that the model dependence of the \emph{shape} of the ratio p/Fe is small (while the absolute value varies significantly). This means that one can safely study differences in the shapes of different spectra in the source against systematic uncertainties of the propagation model.  The greater the difference between the masses of the considered nuclei, the stronger the model dependence, so the latter is the most apparent for p/Fe from all possible combinations in the list of abundant nuclei (p, He, C, O, Ne, Mg, Si, Fe). For all other ratios, the model dependence is weaker than for p/Fe (Figure~\ref{Ratio-p-Fe}). The conclusion is that the shape of the ratios of the reconstructed source spectra of all abundant primary nuclei are only weakly model dependent in the robust region of parameters $\delta_0\in (\delta_r-\Delta\delta,\delta_r + \Delta\delta)$, and $\lambda_0\in (\lambda_r-\Delta\lambda,\lambda_r + \Delta\lambda)$.

\begin{figure}
\includegraphics[width=0.45\textwidth]{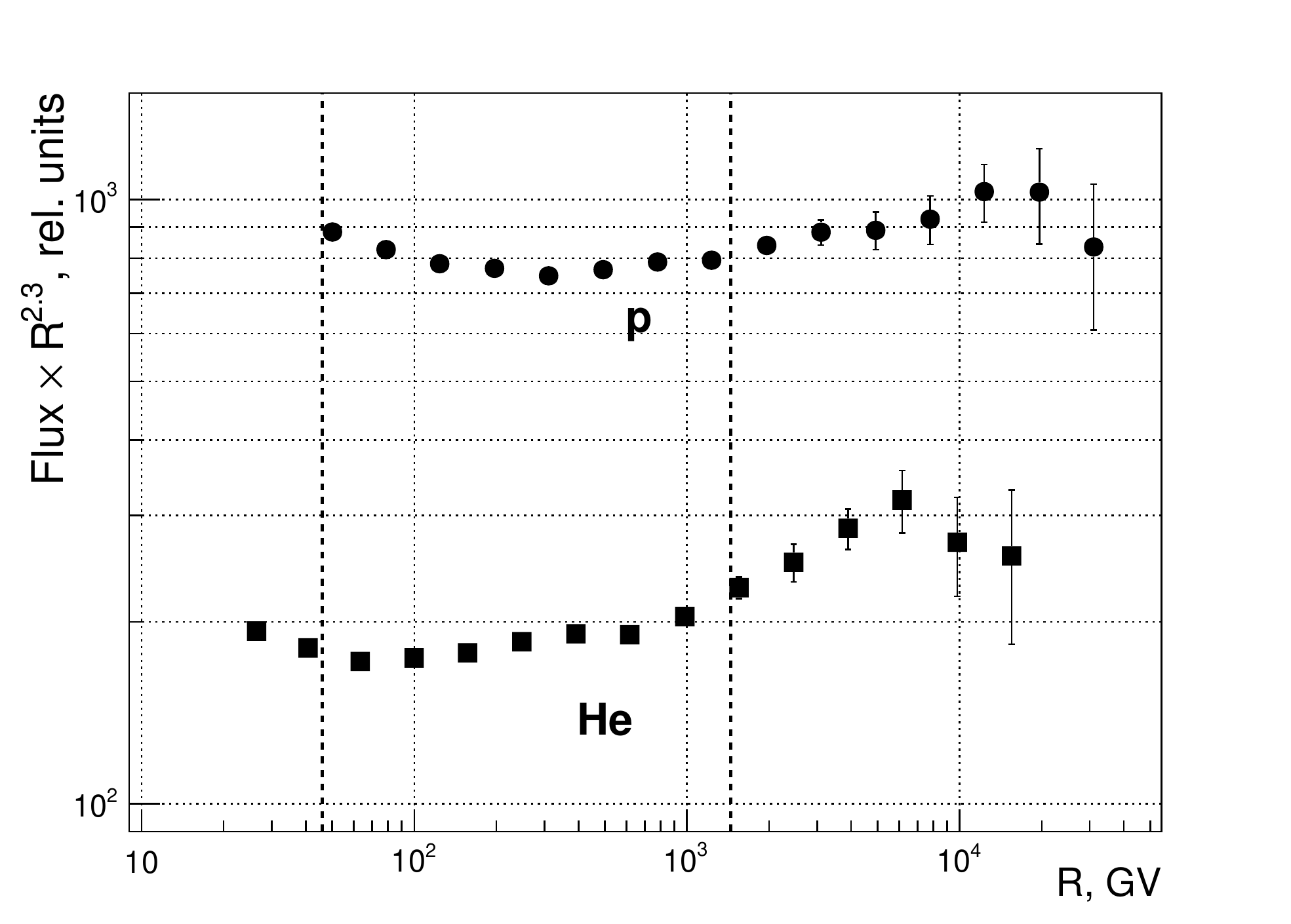}\\
\includegraphics[width=0.45\textwidth]{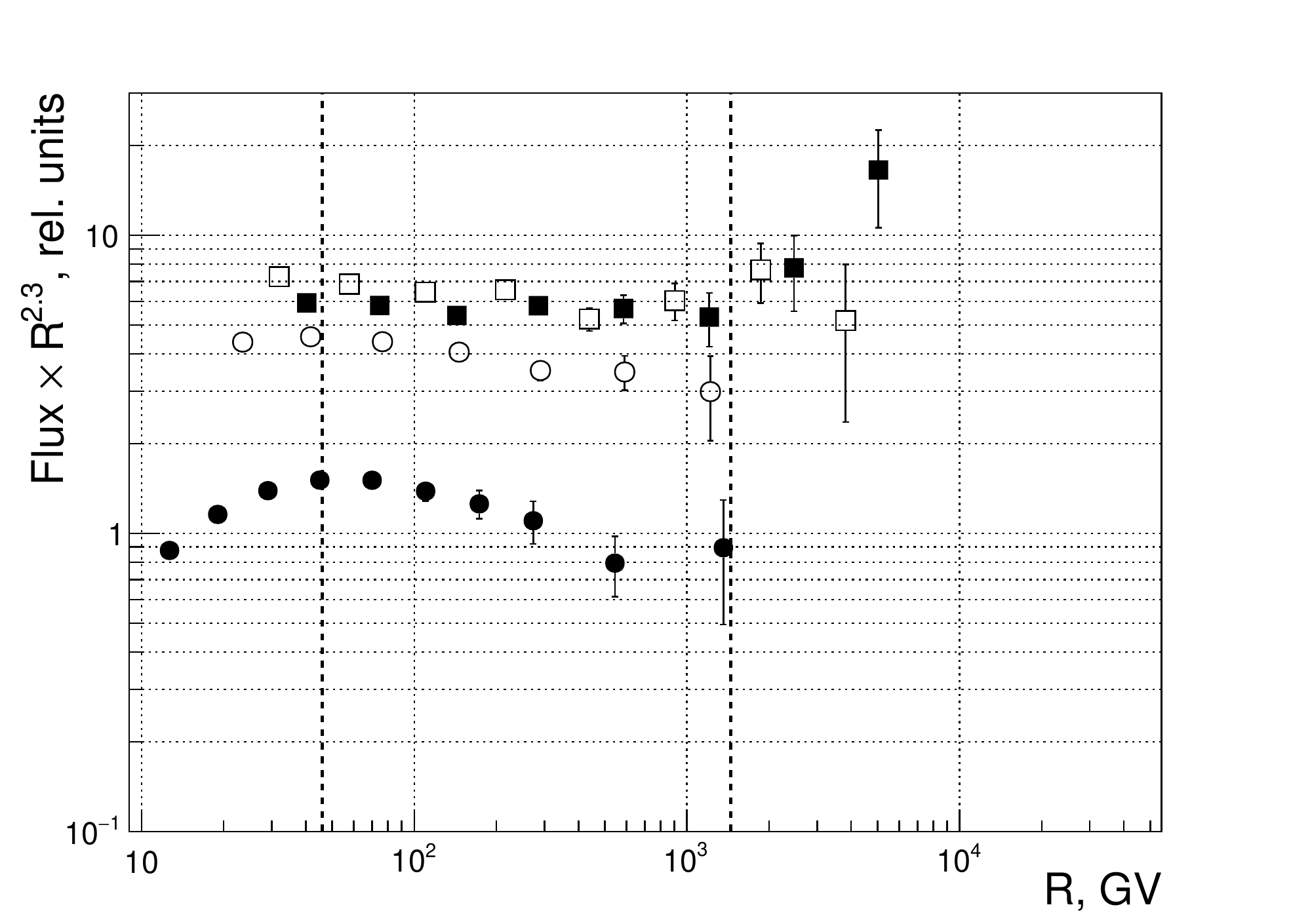}
\caption{\label{AllNucl}Source spectra of nuclei obtained with the reference propagation model (\ref{Ref}).
Top panel: protons and He; bottom panel: C (solid squares), O (open squares), Ne+Mg+Si (open circles), Fe (solid circles).} 
\end{figure}

To improve the statistics, the spectra of nuclei Ne, Mg, and Si are combined for the following into the single rigidity spectrum Ne+Mg+Si with an effective charge number $Z=12$. Figure~\ref{AllNucl} 
shows the source spectra for protons, He, C, O, Ne+Mg+Si, and Fe obtained after the solution of the back-propagation problem with the reference model of Equation~(\ref{Ref}). It can be seen that the spectra of protons and helium are not described by a simple power law, and they became flatter at high energies. This phenomenon in the observed spectra was indicated in the ATIC experiment \citep{ATIC-2009-PANOV-IzvRAN-ENG} and was confirmed by PAMELA \citep{CR-PAMELA-2011-p-He-Mag} and AMS-02 \citep{AMS-02-2015-PRL1,AMS-02-2015-PRL-He}. The spectra of carbon and oxygen also become flatter at high energies. The flattening in the observed spectra was measured in the ATIC experiment \citep{ATIC-2007-PANOV-IzvRAN,ATIC-2009-PANOV-IzvRAN-ENG} for the nuclei heavier than helium and was confirmed by the data of the CREAM experiment \citep{CR-CREAM2010A}. The complicated behavior of the spectra may be related to some nonlinear phenomena in cosmic-ray acceleration \citep{CRA-BEREZHKO-1999}, or the heterogeneous structure of sources and the nearby interstellar medium \citep{CRA-OHIRA2011,ATIC-2011-ZATSEPIN-ICRC}, and may be explained quite naturally by the mixing of two or more sources with different spectra \citep{CR-ZATSEPIN2006}. However, the objective of the present paper is not to discuss the nature and the statistical significance of the complex behavior of the spectra, but to compare the spectra of different nuclei in the source and answer the simple questions: Are the spectra in terms of magnetic rigidity in the source the same for different nuclei or not, and how great are the differences if the shapes of the spectra are not the same?

\begin{figure}
\includegraphics[width=0.45\textwidth]{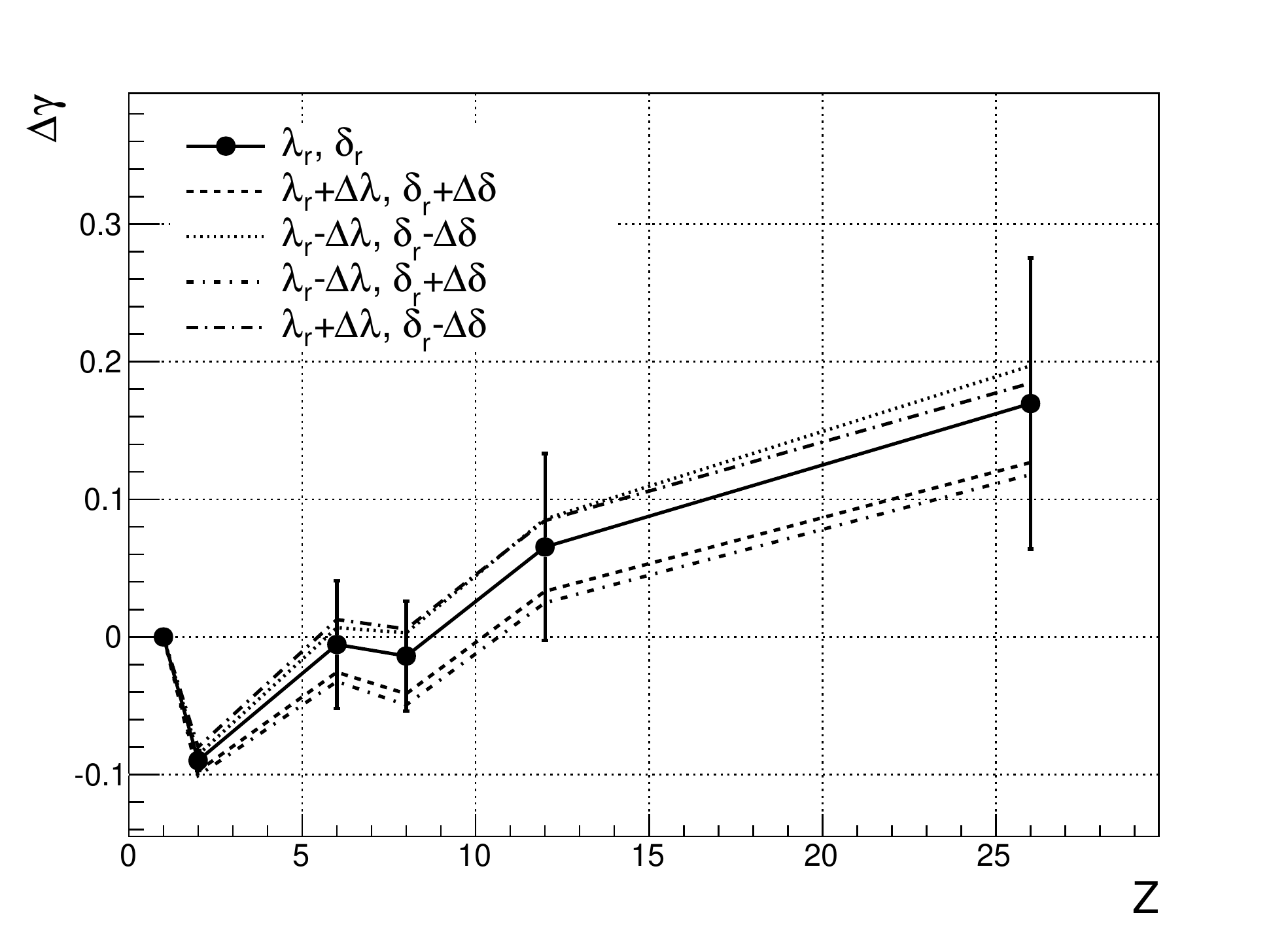}
\caption{\label{DeltaGamma}Differences between the source spectral indices of abundant nuclei and the spectral index of protons obtained using the reference propagation model (\ref{Ref}) and using disturbed parameters for the escape length instead of the reference parameters $\lambda_r$ and $\delta_r$. Statistical errors are specified only for the reference model since for all other models they are the same for each rigidity value $R$.} 
\includegraphics[width=0.45\textwidth]{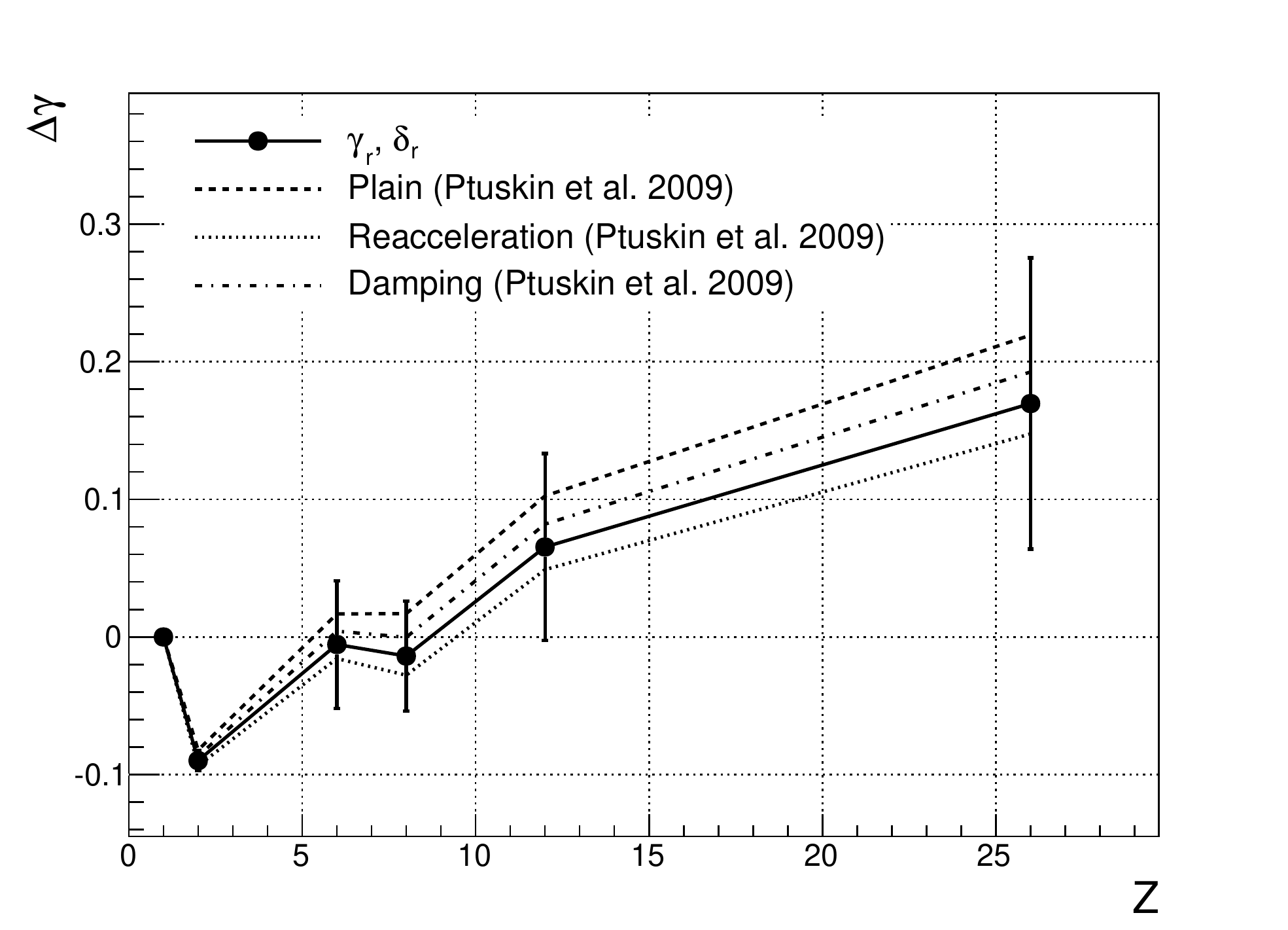}
\caption{\label{DeltaGammaOriginals}Differences between the source spectral indices of abundant nuclei and the spectral index of protons obtained using the reference propagation model (\ref{Ref}) and original propagation models (\ref{PtusL0deltaPlain}--\ref{PtusL0deltaDamping}) of \citep{CRPROP-PTUSKIN2009}.} 
\end{figure}

There are a total of six independent spectra (p, He, C, O, Ne+Mg+Si, Fe), so $6\times(6-1)/2=15$ ratios can be constructed and viewed as approximately model-independent characteristics of cosmic-ray sources. However, 15 ratios are of little use as material for analysis, so in this work we used a simplified approach to grasp important aspects of the behavior of the entire set of the spectra. We took the range of magnetic rigidity common to all of the obtained spectra (approximately 50--1350~GV, indicated by vertical dashed lines in Fig.~\ref{AllNucl}), and we found an average spectral index for each spectrum in this range by approximating it by a power-law function, \emph{ignoring certain deviations from pure power-law behavior} mentioned above. Since the spectral indices were highly model dependent, we were interested not in the spectral indices themselves, but rather in how they varied from one nucleus to another. Figure~\ref{DeltaGamma} 
shows the differences $\Delta\gamma$ between the source spectral indices of abundant nuclei from the spectral index of protons as a function of the nuclear charge. The reconstructed differences between the source spectral indices for disturbed propagation models (as in Figure~\ref{Ratio-p-Fe}) also are shown in Figure~\ref{DeltaGamma} to study the model dependence of the result (or, alternatively, to estimate the systematic errors). It is seen that the model dependence is relatively small, such that the systematic errors are substantially less than the statistical errors in spectral index differences, even for our robust estimate of the systematic errors.

% \begin{figure}
% \includegraphics[width=0.45\textwidth]{DeltaGammaOriginals.pdf}
% \caption{\label{DeltaGammaOriginals}Differences between the source spectral indices of abundant nuclei and the spectral index of protons obtained using the reference propagation model (\ref{Ref}) and original propagation models (\ref{PtusL0deltaPlain}--\ref{PtusL0deltaDamping}) of \citep{CRPROP-PTUSKIN2009}.} 
% \end{figure}

Note that even the original propagation models of Equations (\ref{PtusL0deltaPlain}--\ref{PtusL0deltaDamping}) of \citep{CRPROP-PTUSKIN2009} that were not optimized to the AMS-02 B/C data \citep{AMS-02-2016-PRL-BtoC} do not produce strong systematic deviations in $\Delta\gamma$ from our reference model, as can be seen in Figure~\ref{DeltaGammaOriginals}. 

Discussing the data in Fig.~\ref{DeltaGamma}, it is worth mentioning first that the spectral index of protons \emph{in the source} statistically significantly differs from that of helium: $\Delta\gamma = 0.090\pm0.007(\mathrm{stat})\pm0.011(\mathrm{syst})$. Secondly, the steady rise in the spectrum steepness moving from helium to iron may be noted. This result is also statistically significant, as the slope of this part of the curve in Fig.~\ref{DeltaGamma} is positive with a statistical significance greater than $3.2\sigma$ (both statistics and systematics included). It is not clear, however, whether it makes physical sense to describe all nuclei from helium to iron using one curve, since the helium and heavy nuclei could originate from fundamentally different cosmic-ray sources. It is therefore logical to consider nuclei heavier than helium separately. There is also a positive trend in the slopes of their curves, but it is maintained with a statistical significance of just $1.5\sigma$ (statistics and systematics). We may not therefore speak about an indication of a trend of the spectral index from carbon to iron in the ATIC data.

\begin{figure}
\includegraphics[width=0.45\textwidth]{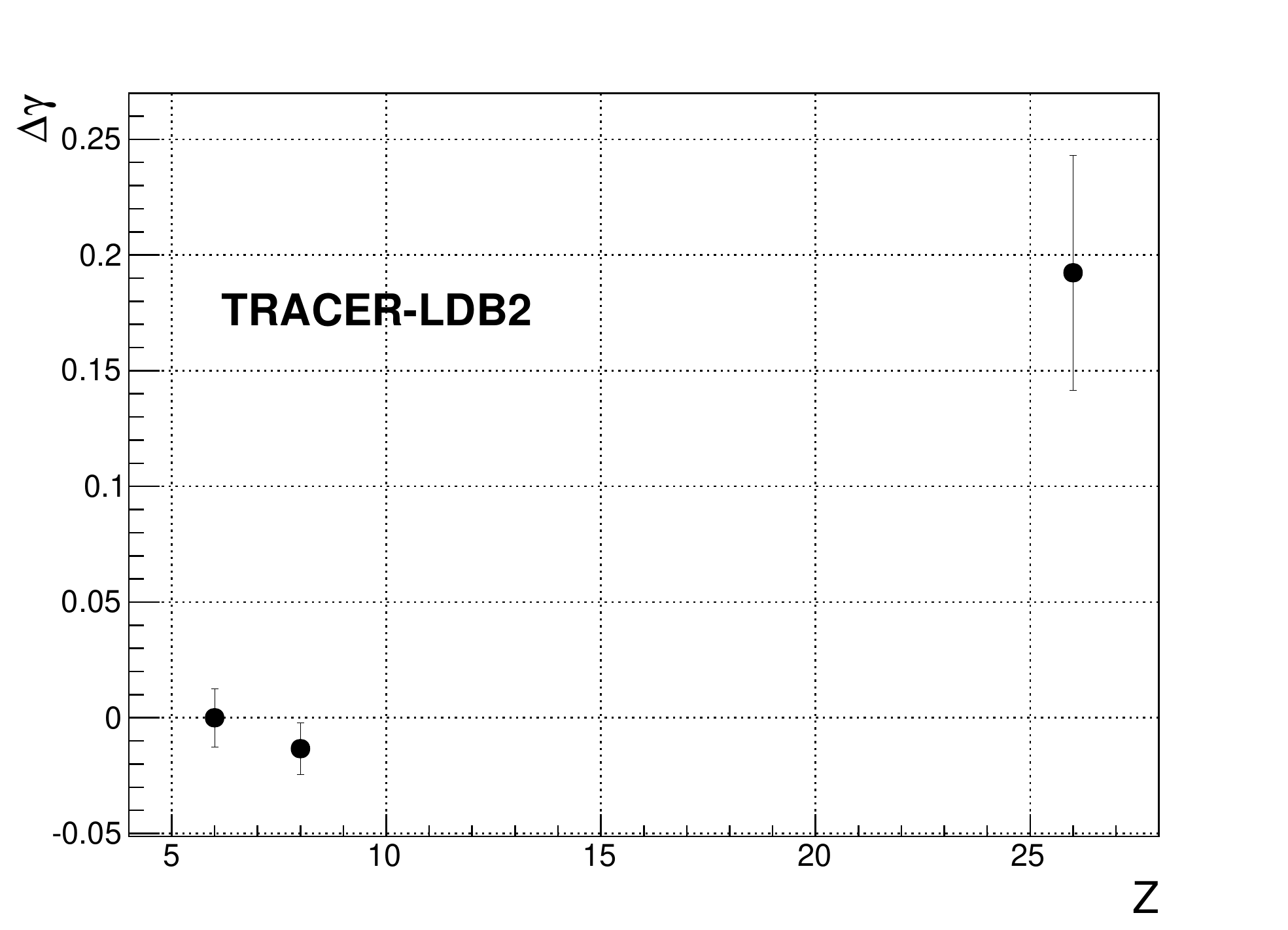}
\includegraphics[width=0.45\textwidth]{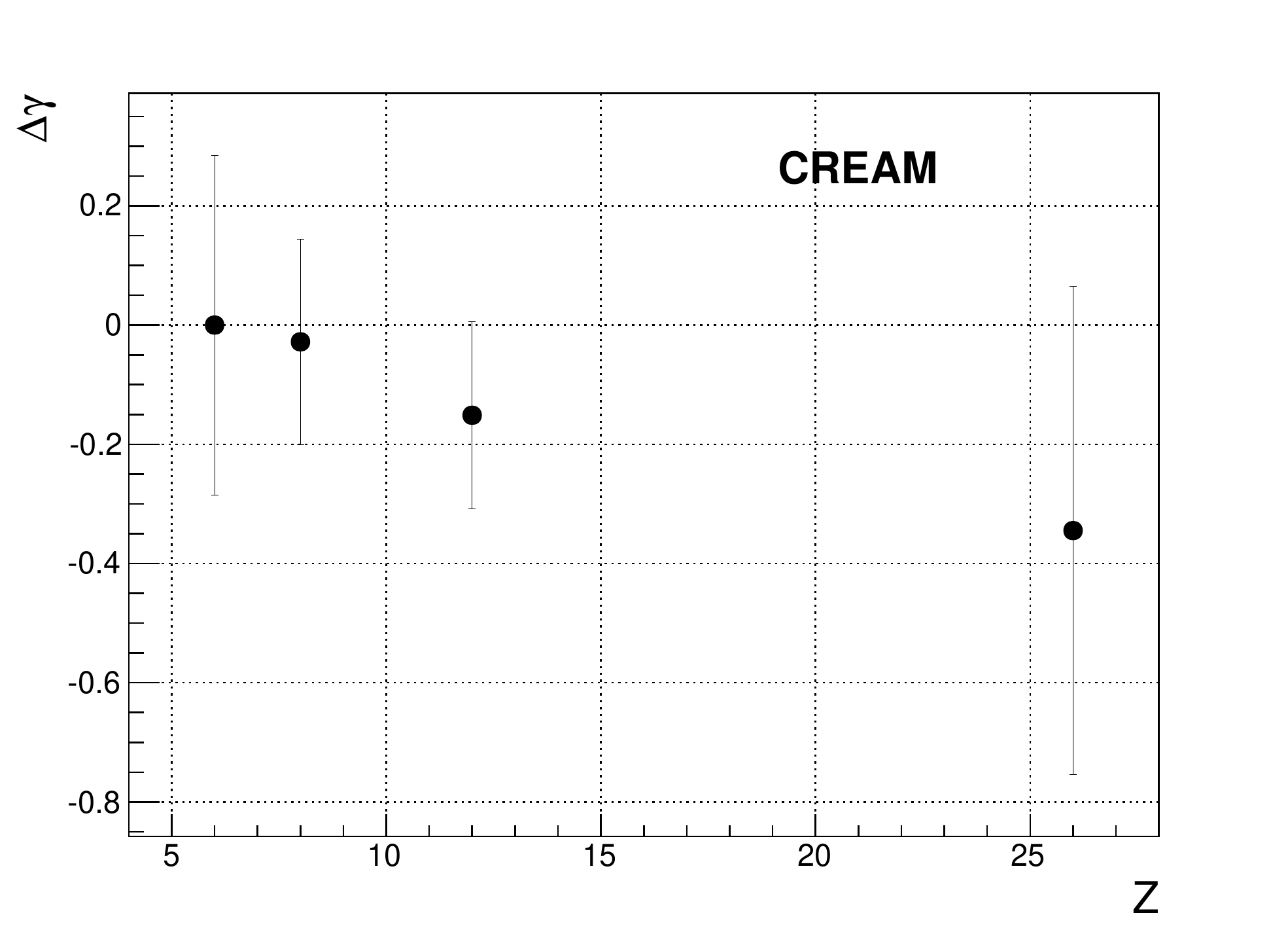}
\caption{\label{OtherExp}Differences between the source spectral indices of C, O, and Fe obtained from the data of the TRACER-LDB2 experiment \citep{CRNUCL-TRACER2011-ApJ} (top panel) 
and CREAM-II \citep{CR-CREAM2010A} (bottom panel). The point for Ne+Mg+Si is also shown for CREAM.}
\end{figure}

There are only a few experiments that can be compared with the results of the present paper. Differences between the source spectral indices of oxygen, iron, and carbon obtained from the data of the TRACER-LDB2 experiment \citep{CRNUCL-TRACER2011-ApJ} and CREAM-II \citep{CR-CREAM2010A} are shown in Fig.~\ref{OtherExp}. The reference point is the spectral index of carbon (separately for TRACER and for CREAM). To obtain these plots, the original data of TRACER and CREAM for measured absolute energy spectra of C, O and Fe were processed by us with the solution of the back-propagation problem as described above. We emphasize that the back-propagation problem has not been solved in the original papers of TRACER  \citep{CRNUCL-TRACER2011-ApJ} and CREAM \cite{CR-CREAM2010A}. The reference model of Equation~(\ref{Ref}) was used to generate the plots in Fig.~\ref{OtherExp}. It is seen that the TRACER data show a more steep spectrum for iron than for carbon and oxygen; this trend confirms the ATIC result (Fig.~\ref{DeltaGamma}). The difference in spectral indices between C and Fe in the TRACER experiment is positive with a statistical significance of $3.5\sigma$. The TRACER-LDB1 data \citep{CRNUCL-TRACER2008B-ApJ} show approximately the same result, but with lower statistical significance. The CREAM data show no trends in spectral indices, but the statistical errors are large (about 0.4 for iron versus an expected difference of spectral indices of 0.2, as may be deduced from the ATIC and TRACER data); therefore, no conclusions may be drawn. We also should note that the results for TRACER and CREAM were obtained for magnetic rigidities less than 400~GV, versus 1350~GV in ATIC. The absolute spectra are not quite power law in all experiments; therefore the results of this comparison should be accepted with caution. The results are related to mean spectral indices in a power-law approximation only, and the energy ranges are similar but not exactly the same for different experiments. Obviously, more exact experimental data are needed to draw more accurate conclusions.

\section{Summary and discussion}
\label{5}

We would like to note that only a rather restricted subset of possible propagation models (homogeneous galaxy halo) was studied in this paper. Our conclusion is that, within this subset of models, the results on ratios of source spectra of different nuclei are almost model independent, but clearly other more complicated propagation models should be studied. We represented our results as differences of averaged source spectral indices of different nuclei in the magnetic rigidity range 50--1350~GV. This approximate method is adequate for rather low statistics and a relatively narrow energy range, for comparison of spectra of different nuclei. New and more precise experiments are needed to obtain and study more detailed information. 

The difference of slopes of observed proton and helium spectra has been considered previously as an indication of different acceleration conditions for protons and helium in the sources many times, starting with the 2004 paper of ATIC \citep{ATIC-2004-ZATSEPIN-IzvRan}, where this difference was observed with high statistical confidence (about $13\sigma$) for the first time. This paper generalizes this important result to all abundant primary nuclei up to iron.  The obtained differences of the slopes of the spectra in the source are a clear indication that the acceleration conditions may vary for all nuclei from protons to iron. This indication is important for understanding the astrophysical mechanisms of the acceleration of cosmic rays, but our opinion is that it is too early to discuss the details of the physics of these variations since the only experiment (ATIC) have reported this phenomenon explicitly. The results should be confirmed by new and more precise experiments.

This work was supported by the Russian Foundation for Basic Research, project no. \mbox{14-02-00919}.

% \bibliography{AticSourceAbundant}
% \bibliographystyle{aasjournal}

\end{document}